\documentclass[sigconf]{acmart} 
\usepackage{graphicx} 
\usepackage{multirow}%
\usepackage{amsthm}%
\usepackage{mathrsfs}%
\usepackage[title]{appendix}%
\usepackage{xcolor}%
\usepackage{textcomp}%
\usepackage{manyfoot}%
\usepackage{booktabs}%
\usepackage{algorithm}%
\usepackage{algorithmicx}%
\usepackage{algpseudocode}%
\usepackage{listings}%
\usepackage[inline]{enumitem}
\usepackage{subcaption}
\usepackage{mathtools}
\usepackage{footmisc}
\usepackage{caption}
\usepackage{comment}
\usepackage{balance}
\definecolor{boxgrey}{HTML}{F3F3F3}

\AtBeginDocument{%
  \providecommand\BibTeX{{%
    \normalfont B\kern-0.5em{\scshape i\kern-0.25em b}\kern-0.8em\TeX}}}

\def\threedigits#1{\number#1}

\newcommand{\hlbox}[2]{%
  \begin{center}
    \fcolorbox{white}{boxgrey}{%
      \parbox{0.9\columnwidth}{\textbf{#1}. \textit{#2}}%
    }
  \end{center}
}

\copyrightyear{2024}
\acmYear{2024}
\setcopyright{rightsretained}
\acmConference[RecSys '24]{Eighteenth ACM Conference on Recommender
Systems}{October 14--18, 2024}{Bari, Italy}
\acmBooktitle{Eighteenth ACM Conference on Recommender Systems (RecSys
'23), October 14--18, 2024, Bari,
Italy}
\acmDOI{XXXXX}
\acmISBN{YYYY}
\begin{document}

\title{Faithful Path Language Modeling for Explainable Recommendation over Knowledge Graph}

\author{Giacomo Balloccu}
\email{gballoccu@acm.org}
\orcid{0000-0002-6857-7709}
\affiliation{%
  \institution{University of Cagliari}
  \streetaddress{Via Ospedale 70}
  \city{Cagliari}
  \country{Italy}
  \postcode{09144}
}

\author{Ludovico Boratto}
\email{ludovico.boratto@acm.org}
\orcid{0000-0002-6053-3015}
\affiliation{
  \institution{University of Cagliari}
  \streetaddress{Via Ospedale 70}
  \city{Cagliari}
  \country{Italy}}

\author{Christian Cancedda}
\email{christian.cancedda@polito.it}
\orcid{0000-0002-6857-7709}
\affiliation{%
  \institution{Polytechnic University of Turin}
  \streetaddress{Corso Castelfidardo, 39}
  \city{Turin}
  \country{Italy}
}

\author{Gianni Fenu}
\email{fenu@unica.it}
\orcid{0000-0003-4668-2476}
\affiliation{
  \institution{University of Cagliari}
  \streetaddress{}
  \city{Cagliari}
  \country{Italy}}

\author{Mirko Marras}
\email{mirko.marras@acm.org}
\orcid{0000-0003-1989-6057}
\affiliation{
  \institution{University of Cagliari}
  \streetaddress{}
  \city{Cagliari}
  \country{Italy}}

\begin{abstract}
The integration of path reasoning with language modeling in recommender systems has shown promise for enhancing explainability but often struggles with the authenticity of the explanations provided. Traditional models modify their architecture to produce entities and relations alternately—for example, employing separate heads for each in the model—which does not ensure the authenticity of paths reflective of actual Knowledge Graph (KG) connections. This misalignment can lead to user distrust due to the generation of corrupted paths. Addressing this, we introduce PEARLM (Path-based Explainable-Accurate Recommender based on Language Modelling), which innovates with a Knowledge Graph Constraint Decoding (KGCD) mechanism. This mechanism ensures zero incidence of corrupted paths by enforcing adherence to valid KG connections at the decoding level, agnostic of the underlying model architecture. By integrating direct token embedding learning from KG paths, PEARLM not only guarantees the generation of plausible and verifiable explanations but also highly enhances recommendation accuracy. We validate the effectiveness of our approach through a rigorous empirical assessment, employing a newly proposed metric that quantifies the integrity of explanation paths. Our results demonstrate a significant improvement over existing methods, effectively eliminating the generation of inaccurate paths and advancing the state-of-the-art in explainable recommender systems. Source code and data: \url{https://tinyurl.com/pearlmrecsys}.
\end{abstract}

\maketitle

\section{Introduction}

\vspace{1mm} \noindent{\bf Motivation.}
The convergence of disciplines such as law, economy, and psychology has spotlighted the critical need for transparency in automated recommender systems (RSs). Notably, regulatory frameworks like the General Data Protection Regulation (GDPR) \cite{GoodmanF17} underscore the 'right to explanation', placing transparency at the forefront of global regulatory and commercial agendas. From a business perspective, transparency not only fosters trust but also enhances user engagement and retention \cite{Tintarev2007}. Psychologically, clear and understandable explanations can trigger curiosity and facilitate decision-making processes \cite{MILLER20191}. However, the opaqueness of traditional RSs, which often operate as black boxes, has prompted a shift towards developing more transparent systems \cite{INR-066}.

\begin{figure}[!t]
\centering
\includegraphics[width=.95\linewidth]{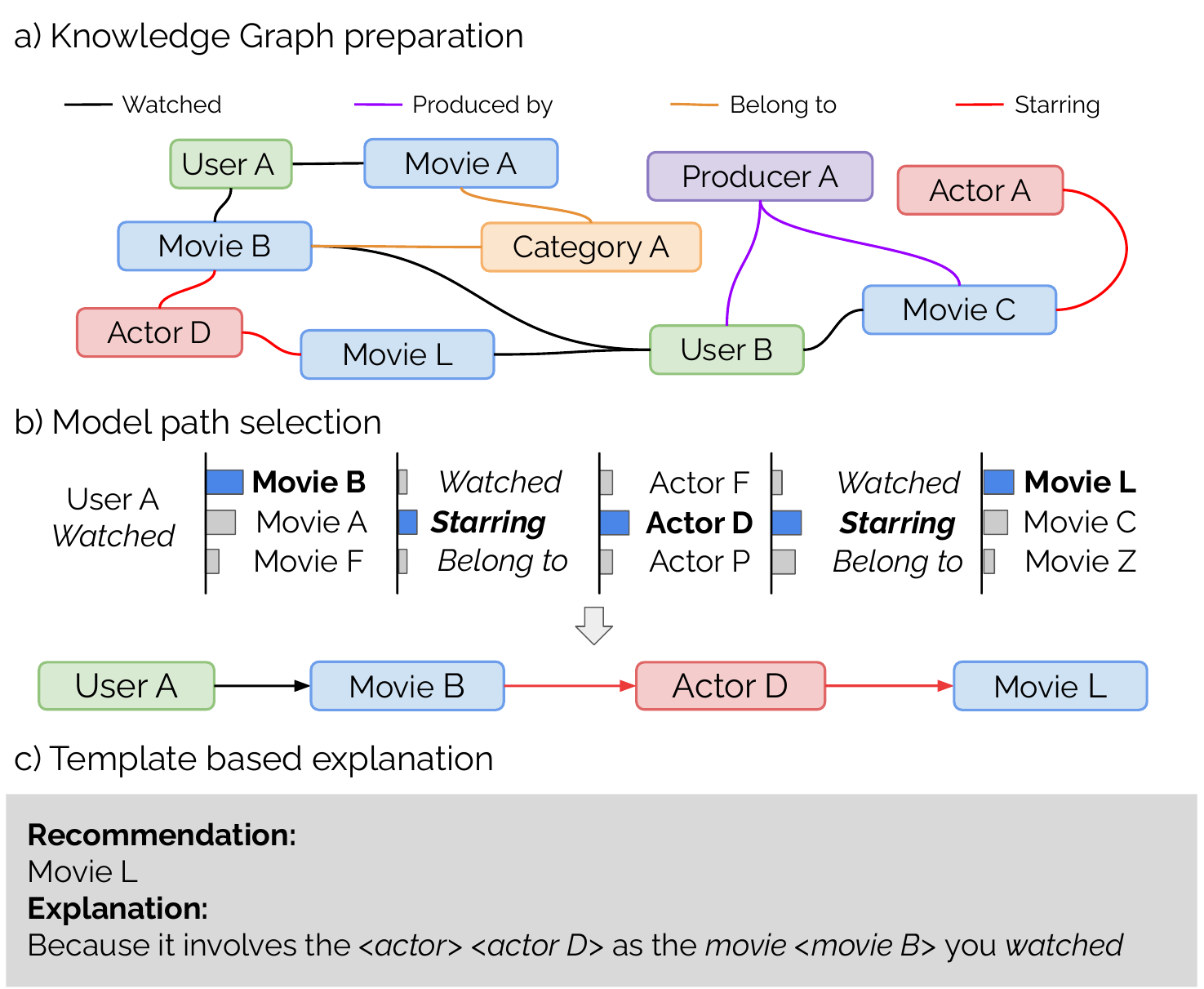}
\vspace{-2mm}
\caption{Given a knowledge graph (a), path-reasoning methods select a path in the knowledge graph that, from a certain user, leads to the product to recommend (b). The selected path is then used for a textual explanation generation step, performed through pre-defined explanation templates (c).} 
\label{fig:introductionary-img}
\end{figure}

\vspace{1mm} \noindent{\bf State of the art.}
Modern recommender systems enhance transparency by coupling recommendations with textual explanations that elucidate the reasoning behind each suggestion. These explanations are derived from either predefined templates or generative methods and utilize user-product relationships encapsulated within a knowledge graph (KG) \cite{TIDDI2022103627}. For instance, a predefined template might generate an explanation like "Because it involves the actress Kate Hudson as in the movie Mother's Day you watched," based on the user's previous interactions, as illustrated in Figure \ref{fig:introductionary-img}.

Traditionally, path selection for explanations occurs post recommendation, employing either human-annotated path types or entity similarity metrics \cite{Ai_2018, wang2018explainable}. However, such post-hoc explanations frequently misalign with the model's actual decision-making process, lack scalability, and potentially misrepresent the model's logic. To overcome these limitations, contemporary approaches integrate recommendation generation with path selection via reasoning techniques employing reinforcement learning (RL). These techniques optimize agents to discover informative paths within KGs, connecting users to relevant but unseen products \cite{10.1145/3331184.3331203, 10.1145/3340531.3412038, usercentric, liu2021reinforced, SAEBI202212, BALLOCCU2023110098}.

Despite advances, RL-based path reasoning methods confront significant challenges, including reliance on pre-trained KG embeddings and managing the balance between exploration and exploitation in sparse KGs. This often leads to lower recommendation quality in terms of utility, coverage, and novelty \cite{10.1007/978-3-031-28241-6_1}.

Recent advancements point to a promising direction with the integration of path reasoning and causal language modeling. These methodologies, capitalizing on the structural similarities between KG paths and natural language, employ Transformer-based models \cite{radford2019language, touvron2023llama}. Initiating with a user entity as the prompt, these models generate sequences alternating between entities and relations in the KG \cite{10.1145/3485447.3511937}, structurally mirroring KG paths and thus offering simultaneous recommendations and explanations.

\vspace{1mm}
\noindent{\bf Open issues.}
The Path Language Modeling (PLM) framework \cite{10.1145/3485447.3511937} currently stands as the only approach utilizing causal language modeling for explainable recommendations over knowledge graphs (KGs). Despite its innovative approach, PLM grapples with critical limitations. One significant issue is the failure to effectively anchor explanations in the factual realities of the KG, undermining the framework's core goal of providing credible explanations for recommended items. Moreover, such method relies on being initialized with pre-trained KG embeddings, learned separately and with different objectives, whose limited size may also constrain the potential of advanced language models \cite{radford2019language}. Specifically, \cite{10.1145/3485447.3511937} employs separate parameter sets (heads) for entities and relations within the KG, an attempt to make the model adhere to the KG's entity-relation structure at the model level. While structuring outputs to reflect entity-relation sequences, this approach still generates incorrect triplets and leads to increased computational complexity and entities-relations independence assumptions.

A primary concern is the phenomenon of \emph{model hallucination in path reasoning} (Section~\ref{sec:model-hallucination}), where models produce explanations based on non-existent KG relations. This leads to severe misalignments, compromising the integrity and credibility of the provided explanations. Our detailed analysis using the MovieLens and LastFM datasets (Section~\ref{subsec:measuring-faithfulness}) demonstrates the necessity for models that generate more accurate and KG-faithful paths. Further insights from a comprehensive user study (Section~\ref{subsec:user-survey}) highlight how such inaccuracies impact user trust and perception. These findings show the critical need for methodologies that ensure the generation of plausible and structurally coherent paths, thereby maintaining the reliability and transparency of explainable recommender systems.

\vspace{1mm} \noindent{\bf Our approach, contributions.} 
In response to these challenges, we introduce PEARLM (Path-based Explainable-Accurate Recommender based on Language Modelling), a novel methodology in the domain of path reasoning with causal language modelling (CLM). PEARLM's uniqueness stems from three main innovations: i) the introduction of \emph{KG-constrained sequence decoding} to ensure the authenticity of the generated paths, (ii) a unified model architecture for modelling entities and relations jointly, and (iii) learning of token embeddings specialised for path language modelling from KG paths, bypassing the need for pre-trained embeddings optimised solely for link prediction tasks \cite{NIPS2013_1cecc7a7}. Our contributions include:
\begin{itemize}[leftmargin=5mm]
\item A metric to assess and empirical study on hallucination in KG-based explainable recommendation systems, emphasising its effects on users' perception of explanations and the challenges experienced by users in detecting such inaccuracies (Section~\ref{sec:model-hallucination}).
\item A high-performing path reasoning method using causal language modelling, enhanced by KG-constrained sequence decoding, which inherently produces accurate entity and relation predictions grounded in the KG (Section~\ref{sec:tecnical-details}).
\item Extensive experiments across varied datasets, demonstrating PEARLM's efficacy in enhancing recommendation utility and beyond accuracy goals (Section~\ref{sec:rq1}-\ref{sec:rq2}).
\item An analysis of key modelling factors, such as direct embedding learning, the use of KGCD, as well as sampled path properties and model parameter size dataset size, confirming PEARLM's effectiveness and disentangling the effects of each component on model performance (Section~\ref{sec:rq3}).
\end{itemize}

\section{Preliminaries}
This section introduces preliminary concepts crucial for understanding KG-based explainable recommendation systems.

\noindent{\bf Knowledge Graph (KG).} A Knowledge Graph is represented by a set of triplets \(\mathcal{G} = \{(e_h, r, e_t) | e_h, e_t \in \mathcal{E}, r \in \mathcal{R}\}\), where \(\mathcal{E}\) and \(\mathcal{R}\) denote the sets of entities and relations, respectively. Each triplet \((e_h, r, e_t)\) describes a relationship \(r\) between a head entity \(e_h\) and a tail entity \(e_t\). In the context of explainable recommendations, entities include users (\(\mathcal{U} \subset \mathcal{E}\)) and products (\(\mathcal{P} \subset \mathcal{E}\)), and relationships might represent interactions (\(r_f\)) such as 'watched' or 'purchased'. Entities like movie actors or music producers are linked via relations reflecting their roles (e.g., 'acted in', 'produced').

\noindent{\bf K-Hop Reasoning Path.} A \(k\)-hop path in \(\mathcal{G}\) between entities \(e_0\) and \(e_k\) is a sequence of entities connected by \(k\) relations, \(l^k_{e_0, e_k} = \{ e_0 \xleftrightarrow[]{r_1} e_1 \xleftrightarrow[]{r_2} \dots \xleftrightarrow[]{r_k} e_k \}\), where each pair \(e_{i-1} \xleftrightarrow[]{r_i} e_i\) corresponds to a triplet or its inverse in \(\mathcal{G}\). We denote as $\mathcal L^k_{u}$ the set of all \emph{user-centric} possible $k$-hop paths between a user $u$ and any product. We distinguish between user-centric paths leading to previously interacted products (\(\mathcal{\grave{L}}^k_{u}\)) and potential new interactions (\(\mathcal{\tilde{L}}^k_{u}\)).

\noindent{\bf Recommendation over Knowledge Graphs.} The matrix \(\mathcal F \in \mathbb R^{|\mathcal U| \times |\mathcal P|}\) represents user-product interactions, with \(\mathcal F(u, p) = 1\) indicating an interaction. Traditional models estimate the relevance \(\Tilde{\mathcal F}(u, p) \in [0, 1]\) for unobserved interactions to rank products, recommending the top-\(n\) products \(\Tilde{\mathcal P}^n_u\) to user \(u\). To improve transparency, recommendations are associated with a selected \(k\)-hop path, \(\Tilde{l}^k_{u,\tilde{p}}\), enhancing the explanation of recommended products.

\noindent{\bf Autoregressive Path Generation for Explainable Recommendation.} We develop a model \(\theta\) for generating paths within \(\mathcal{G}\), aiming to recommend products through user prompts starting with \(u, r_f\) and adhering to KG structural constraints. The model predicts the next token \(t\) (entity or relation) in a path \(l^{k-1}_{u,p}\), formalized as \(\theta : l^{k-1}_{u,p} \rightarrow \mathbb{P}(V | l^{k-1}_{u,p})\), where \(V\) includes all entities and relations. During the generation, the model iteratively selects tokens to construct meaningful paths that lead to novel product recommendations, scoring each path to present the most relevant products $\Tilde{\mathcal P}^n_u$ along with their explanations paths $\Tilde{l}^k_{u,p}$ to the user.

\section{Exploratory Analysis}
\label{sec:model-hallucination}
Model hallucination in path reasoning presents a significant challenge in explainable recommendations, occurring when models generate paths that inaccurately represent the underlying Knowledge Graph (KG). Specifically, models may erroneously establish incoherent semantic relations between entities, such as extending user-item connections with inappropriate relations like "starred by" which should logically apply only between an actor and a movie within the KG. Moreover, incoherence can arise when entities semantically connected in the real world are linked in the KG through non-existent relationships, such as incorrectly linking "Johnny Depp" with the relation "starred in" to the non-existent KG item "Interstellar."

These inaccuracies undermine the fundamental rationale for utilizing a KG by misaligning the factual truths that should underpin the explanations offered. This section formalizes a metric to assess the extent of path corruption produced by path reasoning methods, presents empirical results from the only existing approach, and analyzes the impacts of this hallucination through a user study.

\subsection{Path Faithfulness Operationalization}
\label{subsec:measuring-faithfulness}
To rigorously evaluate path reasoning models, we define the Path Faithfulness Rate at hop \(k\) (PFR@k) within the structure of a Knowledge Graph (KG). Let \(\mathcal{\tilde{L}}^k_{u,p}\) be the set of all \(k\)-hop paths generated by the model from a user \(u\) to any product, and let $\mathcal{\tilde{L^*}}^k_{u,p}$ denote the subset of these paths that do not contain any corrupted triplets up to the \(k\)-th hop. A path is considered corrupted if any of its triplets \( (e_h, r, e_t) \) from the sequence is ill-formed or not present in the KG, as defined by \( \mathcal{G} \). Thus, PFR@k is mathematically formulated as:
\[
\text{PFR@k} = \frac{|\mathcal{\tilde{L^*}}^k_{u,p}|}{|\mathcal{\tilde{L}}^k_{u,p}|}
\]
where \(|\mathcal{\tilde{L^*}}^k_{u,p}|\) is the cardinality of the set of uncorrupted paths up to the \(k\)-th hop, and $|\mathcal{\tilde{L}}^k_{u,p}|$ is the total number of paths generated. This metric quantifies the accuracy and reliability of the paths generated by the model, reflecting their adherence to the factual structure of the KG up to the specified hop \(k\).

\begin{table}[!t]
\centering
\caption{PFR@k at various hop size for PLM \cite{10.1007/978-3-031-28241-6_1}}
\label{tab:motivation-example}
\resizebox{0.9\linewidth}{!}{
\begin{tabular}{|l|c|c|c|}
\hline
\textbf{Dataset} & \textbf{PFR@1} & \textbf{PFR@2} & \textbf{PFR@3} \\
\hline
\text{MovieLens (ML1M)} & 0.58 & 0.35 & 0.06 \\
\text{LastFM (LFM1M)} & 0.65 & 0.24 & 0.10 \\
\hline
\end{tabular}
}
\end{table}

\vspace{1mm} \noindent \textbf{Empirical Evaluation of Path Faithfulness} To validate the Path Faithfulness Rate at hop \(k\) (PFR@k) and to illustrate the limitations of existing PLM \cite{10.1145/3485447.3511937} in producing accurate explanations, we analyzed two widely used datasets, MovieLens and LastFM. These datasets were carefully pre-processed to align with our evaluation setup, allowing for a fair path faithfulness assessments. The summarized results, depicted in Table~\ref{tab:motivation-example}, clearly highlight the degradation in path faithfulness with increased hops, indicating the need for models that better adhere to the KG's factual content. The PFR@k metric can be instrumental not only for assessing model performance but also for guiding the training of machine learning models to enhance their ability to generate accurate and coherent paths. 

\subsection{Exploration on Users' Perceptions}
\label{subsec:user-survey}
Before developing PEARLM, we conducted a user survey\footnote{Survey details available at \href{https://tinyurl.com/survey-copy}{https://tinyurl.com/survey-copy}} to explore how explanation faithfulness influences user perception. This study primarily aimed to determine if inaccuracies such as incorrect or incoherent semantic relations in explanations could negatively impact users' perceptions of explanation quality.

\vspace{1mm} \textbf{Survey Design and Hypotheses.} Our survey was structured to test two primary hypotheses: (i) users' awareness of potential inaccuracies reduces their perceived quality of explanations, and (ii) inaccuracies within explanations are challenging for users to detect, irrespective of prior warnings. We employed established metrics \cite{4401070} like Transparency, Effectiveness, Efficiency, Satisfaction, Scrutability, Persuasiveness, and Trust, measured on a 1-3 Likert scale \cite{10.1145/3397271.3401032}, to assess user reactions to different types of explanations, including baseline, collaborative, and path reasoning (both accurate and containing inaccuracies). Each participant evaluated multiple explanations, prompting them to report any detected inaccuracies.

\vspace{1mm} \noindent \textbf{Explanation Types and User Profiles.}
Participants were shown explanations for four random recommendations from a pool of eight, ensuring each user evaluated two instances of each explanation type. This strategy was employed to reduce individual bias towards specific recommendations and ensure a balanced exposure to various explanation styles. Each recommendation was accompanied by two explanations, with in total 16 evaluations per user. The list comprised a mix of baseline explanations, collaborative explanations, and path reasoning explanations. Specifically, given the recommendation "You may like \textit{Recommended Movie}...", we compared three types of explanations: (i) baseline explanations (BL) commonly used in production environments, such as "... because you watched $movie_1$," (ii) collaborative explanations (COL) akin to  "... based on a similar user who also watched $movie_1$," and (iii) path reasoning explanations (PR) like \emph{"... because it shares $entity\_type$ $entity\_name$ with $movie_1$, a movie that you previously enjoyed."} Explanations were composed of two collaborative, two baseline, and four path reasoning explanations. For the latter, two explanations were manually crafted and accurate (PR-A), while other two were produced by PLM and corrupted with inaccuracies (PR-I).


\vspace{1mm} \noindent \textbf{Participant Sampling and Grouping.}
We engaged 164 participants via the Prolific platform, ensuring balanced gender and country representation. These participants were divided into two groups: i) the 'blind group', which was not informed about the potential inaccuracies in the explanations, and ii) the 'informed group', which received a caution about possible errors. Random assignment to these groups was performed, ensuring an unbiased distribution. Additionally, to control for individual preference bias in movie recommendations, each participant was assigned a randomly generated user profile at the beginning of the questionnaire. These profiles were constructed to reflect the interaction history of typical users, and participants were presented with four recommendations randomly chosen from a set of eight. This approach aimed to simulate a realistic user experience while ensuring that evaluations of the explanations were based solely on their quality, rather than personal movie preferences.

\begin{table}[!t]
\centering
\caption{Perception of Explanation Quality}
\label{tab:survey-results}
\resizebox{1\linewidth}{!}{
\begin{tabular}{|l|c|c|c|c|c|c|c|}
\hline
\multicolumn{8}{|c|}{\textbf{Blind Group}} \\
\hline
\textbf{Explanation Type} & \textbf{EFFE} & \textbf{EFF} & \textbf{PERS} & \textbf{SAT} & \textbf{SCR} & \textbf{TRA} & \textbf{TRU} \\
\hline
BL & 1.59 & 1.6 & 1.61 & 1.50 & 1.72 & 1.66 & 1.67 \\
\hline
COL & 1.46 & 1.71 & 1.51 & 1.43 & 1.46 & 1.94 & \underline{1.86} \\
\hline
PR-A & \textbf{1.81} & \textbf{1.95} & \textbf{2.02} & \textbf{1.94} & \textbf{2.00} & \textbf{2.29} & \textbf{2.08} \\
\hline
PR-I & 1.41 & 1.64 & 1.54 & 1.31 & 1.61 & 2.00 & 1.51 \\
\hline
PR-A + PR-I & \underline{1.66} & \underline{1.83} & \underline{1.84} & \underline{1.70} & \underline{1.85} & \underline{2.19} & \underline{1.86} \\
\hline
\end{tabular}
}
\hspace{5mm} 
\resizebox{1\linewidth}{!}{
\begin{tabular}{|l|c|c|c|c|c|c|c|}
\hline
\multicolumn{8}{|c|}{\textbf{Informed Group}} \\
\hline
\textbf{Explanation Type} & \textbf{EFFE} & \textbf{EFF} & \textbf{PERS} & \textbf{SAT} & \textbf{SCR} & \textbf{TRA} & \textbf{TRU} \\
\hline
BL & 1.69 & 1.67 & \textbf{1.94} & 1.78 & 1.68 & 1.98 & \underline{1.65} \\
\hline
COL & 1.56 & 1.74 & \textbf{1.94} & \textbf{1.80} & 1.73 & 2.04 & \textbf{1.72} \\
\hline
PR-A & \textbf{1.78} & \textbf{1.88} & 1.67 & \underline{1.79} & \textbf{1.75} & \textbf{2.16} & \textbf{1.72} \\
\hline
PR-I & 1.76 & 1.68 & \underline{1.72} & 1.76 & 1.73 & 1.98 & 1.39 \\
\hline
PR-A + PR-I & \underline{1.77} & \underline{1.78} & 1.70 & 1.78 & \underline{1.74} & \underline{2.07} & 1.55 \\
\hline
\end{tabular}
}
\vspace{-6mm}
\end{table}

\vspace{1mm} \noindent \textbf{Empirical Study Results.}
The survey revealed crucial insights about how user awareness and explanation type affect perceived explanation quality (refer to Table~\ref{tab:survey-results}).

\textit{Overall Perception of Path Reasoning Explanations.}
Across both informed and blind groups, path reasoning explanations (PR), whether accurate (PR-A) or inaccurate (PR-I), were generally perceived more favorably in terms of Transparency and Trust compared to baseline (BL) and collaborative (COL) explanations. This finding supports the hypothesis that PR can enhance explanation quality by providing more detailed and logically connected narratives.

\textit{Blind Group Results.}
In the blind group, path reasoning explanations that were accurate (PR-A) outperformed the baseline (BL) and collaborative (COL) explanations across several metrics. Even when path reasoning explanations contained inaccuracies (PR-I), they were perceived as more transparent than BL and COL, suggesting an intrinsic perceived transparency for the path reasoning format.

\textit{Informed Group Results.}
If informed of potential inaccuracies, which simulates and enviroment when a user from its historical interaction with the system knows that it can do mistakes, participants rated all explanation types lower on average. Baseline explanations saw drops in Persuasiveness (PERS: 1.94) and Satisfaction (SAT: 1.78) compared to the blind group. Yet, collaborative explanations were rated higher in Transparency (TRA: 2.04) and Trust (TRU: 1.72) than the baseline, indicating that even with potential inaccuracies, users value the social proof element in COL.

\textit{Detection of Inaccuracies.} Our survey showed distinct patterns in users' ability to detect inaccuracies in recommendation system explanations. In the 'blind group', 10.7\% incorrectly identified inaccuracies in accurate explanations (PR-A), compared to 22.0\% in the 'informed group'. Conversely, 33.3\% of the blind group and only 13.3\% of the informed group correctly identified actual inaccuracies in PR-I. This suggests that while informed participants were more sceptical, they were less adept at identifying true inaccuracies. Overall, only 23.3\% of participants across both groups successfully detected inaccuracies, highlighting the difficulty in discerning errors in explanations.

This survey highlights the critical role of explanation accuracy and user awareness in shaping trust and satisfaction within recommendation systems. The findings underscore the necessity for developing methodologies like PEARLM that prioritize accuracy and transparency, thereby enhancing user trust and the effectiveness of explainable systems. These insights motivate the design of PEARLM to produce more reliable explanations, aiming to maintain user confidence and improve system integrity. 

\begin{figure*}[!t]
\centering
\includegraphics[width=1\linewidth]{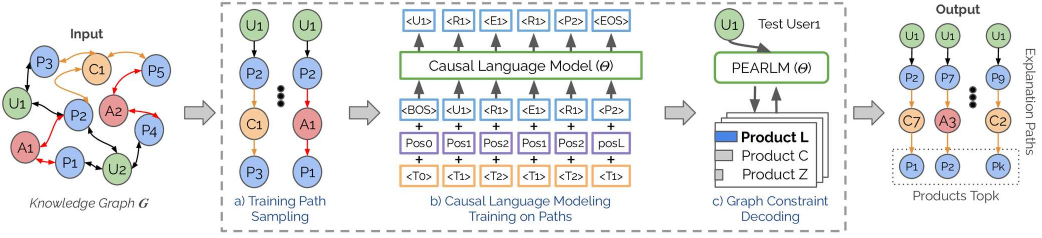}
\caption{PEARLM framework.  The input is represented by the user-product KG $\mathcal G$. This graph is utilised in (a) to extract user-centric paths which are used to pretrain a causal language model (b). With the learned model $\theta$, we apply our decoding in (c) to yield reasoning paths and recommendations.} 
\label{fig:model-pipeline}
\end{figure*}

\section{The Proposed Methodology}
\label{sec:tecnical-details}
In this section, we outline the methodology for constructing training path sequences for our \textbf{Path-based Explainable-Accurate Recommender via Language Modelling} (PEARLM). We then detail our framework (Figure \ref{fig:model-pipeline}), covering the embedding layer, model architecture, and training objectives. Our novel graph constraint decoding technique, designed to extract candidate path sequences for generating faithful explanations, is also expounded.

\subsection{User-Centric Path Construction Learning}
\noindent{\bf Training Path Sampling Procedure.}
Knowledge Graphs (KGs) play a crucial role in revealing complex relationships and patterns. Given the intricate nature of KGs, strategic path sampling is essential for accurately capturing significant connections from user to product. In our approach, we specifically utilize KGs to extract paths that begin with a user interacting with one product and end with a different product interaction. This method is fundamental for capturing and representing user behavior and product-related knowledge, thereby allowing the model to detect and learn from patterns existing between user interactions.

Employing a standard random walk algorithm \cite{li2015random}, we set the desired maximum number of hops, $N \in {3,5}$. Each path initiates with a user entity $u$, followed by an interaction relation $r_f$ and an interacted product $\grave{p}$. The path continues, alternating between entities $e \in E$ and relations $r \in R$, until reaching the predefined hop count $N$, and concludes with another interacted product entity $\grave{p}$. Thus, a valid sequence $S$ is represented as $S = (u, r_f, \grave{p}, r_2, ..., r_{N-1}, \grave{p})$. We sample $\mathcal{K}$ such paths for each user $u$, where $\mathcal{K}$ denotes the sample size. Duplicate paths are excluded to ensure diversity.

\noindent{\bf Vocabulary and Embeddings.}
Considering the duality between paths and sequences of tokens for language modelling, in our work each node and edge in the KG, respectively representing an entity or relation, is defined as a \emph{token} $t$ in a vocabulary encompassing all external entities $\mathcal{E}$, products $\mathcal{P}$, users $\mathcal{U}$ and relations $\mathcal{R}$ in the knowledge graph. 
We use a WordLevel tokenizer to transform entities and relations from the KG into tokens. This results in a dictionary $V$ encompassing all unique tokenized entities and relations $t \in V$.
To capture their semantic meaning, each token is associated with a respective embedding $\mathcal{E}_{V}$.
Besides the vocabulary embeddings, we introduce type embeddings, $\mathcal{E}_{T}$, and positional embeddings, $\mathcal{E}_{P}$. Type embeddings are used to differentiate between entities, relations, and special tokens such as `bos' (beginning of sequence) and `eos' (end of sequence). Positional embeddings are crucial for maintaining the sequence order of tokens as they appear in the sampled paths. Specifically, these embeddings encode the relative positions of entities and relations within a path, preserving the sequential integrity and contextual relationships in the path structure.
The final input embeddings $\mathcal{E}_{F}$ is:
\begin{equation}
\mathcal{E}_{F} = \mathcal{E}_{V} + \mathcal{E}_{T} + \mathcal{E}_{P}
\end{equation}
This composition enables the model to learn the meaningful representations of entities and relations within the KG, while also preserving the sequence structure.
The maximum context length \( L \) that PEARLM can process is determined by the formula \( L = 2 * N + 1 \). This is because each sampled path in a KG comprises an alternating sequence of entities and relations, starting and ending with an entity. For a path with \( N \) hops, there are \( N \) entities and \( N-1 \) relations. The additional '+1' in the formula accounts for the starting entity, thus ensuring the entire path, including all entities and relations, is captured within the model's context.

\noindent{\bf Learning to costruct paths.}
PEARLM innovatively adapts Transformer decoder layers to specifically address the challenges of autoregressive path language modelling in knowledge graphs. This adaptation is key to accurately capturing the sequential nature and complex dependencies inherent in KG paths.

In conventional Transformer models, the attention mechanism broadly assesses the importance of each token relative to others in a sequence. PEARLM's implementation, however, is fine-tuned to emphasise the unique characteristics of KGs. It specifically focuses on the interplay between entities and relations within KG paths, a critical aspect often overlooked in standard language models.

The model's architecture is designed to be particularly sensitive to the sequential flow and hierarchical structure of KG paths. This is achieved through a tailored 'masked' self-attention mechanism, ensuring that the generated predictions adhere to the chronological order and logical consistency of the paths in KGs. This nuanced approach is crucial for maintaining the integrity of path sequences, a requirement for explainable recommendations in KGs.

During training, PEARLM optimises the sequence of embeddings $\mathcal{E}_{F}$ to predict the most probable next token in a KG path. This is realised through a calculated combination of linear transformation and softmax activation at each timestep $k$:
$$
P(t_k | t_1, t_2, ..., t_{k-1}; \theta) = \text{softmax}(W h_k + b)
$$
Where \( W \), \( b \), and \( \theta \) are parameters fine-tuned to enhance the model’s predictive accuracy. By dynamically updating these parameters based on observed paths, PEARLM effectively learns the probabilistic structure of user interactions within the KG, significantly improving its recommendation capability.

This specialized adaptation of Transformer architecture ensures that PEARLM is particularly well-suited for generating explanations based on the actual user interaction paths, enhancing both the accuracy and explainability of recommendations in knowledge-graph-based systems.

\subsection{Knowledge Graph Constraint Decoding}
Decoding in causal language models (CLM), such as GPT-2 \cite{radford2019language}, involves generating a sequence of tokens sequentially. In PEARLM, this process is enhanced with our innovative Knowledge Graph-Constrained Decoding (KGCD) method, ensuring adherence to the structural integrity of the KG. Given a sequence of tokens $t_1, t_2, ..., t_{k-1}$, the CLM predicts the next token $t_k$ based on its predecessors, formalized as $t_k = \text{argmax}_{t \in V} P(t | t_1, t_2, ..., t_{k-1}; \theta)$, where $P$ denotes the probability distribution over the vocabulary, $t$ are potential next tokens, and $\theta$ encapsulates the model parameters.

KGCD in PEARLM diverges from standard decoding processes by incorporating KG constraints directly into sequence generation. In this approach, tokens alternate between 'entity' (e) and 'relation' (r) types, beginning with an entity. The probability of the next token $t_k$ under KGCD is computed as:

$$
P(t_k | t_1, ..., t_{k-1}; \theta) =
\begin{cases} 
\underset{t \in V}{\text{argmax}} P(t | t_1, ..., t_{k-1}; \theta) & \text{if } \psi(t_1,...,t_{k-1}) \\
-\infty & \text{if } \neg \psi(t_1,...,t_{k-1})
\end{cases}
$$

Here, $\psi(t_1,...,t_{k-1})$ is a function determining if a candidate token $t$ is logically reachable in the sequence $t_1,...,t_{k-1}$ based on KG constraints. If $t_k$ is unreachable, its probability is set to negative infinity, effectively preventing its generation. This unique feature of PEARLM ensures that the resulting sequences are not only coherent but also faithfully represent the actual KG structure.

Importantly, KGCD in PEARLM operates at the logit level and is model-agnostic, allowing its integration with other Autoregressive Path Generation models. For each user prompt, PEARLM generates sequences using beam search, divided into beam groups with a diversity penalty to promotes exploration of diverse path options. The generated sequences are then ranked based on the average probability of each token, and the top $k$ sequences are selected. These sequences serve a dual role: they are the primary product recommendations for the user and simultaneously provide the rationale behind each recommendation. 

\section{Experimental Evaluation}
We design our experiments to answer three research questions:

\begin{enumerate}[label={\textbf{RQ\protect\threedigits{\theenumi}}}, leftmargin=*]
\item Does PEARLM enhance recommendation utility metrics compared to  knowledge aware and path reasoning models?
\item How does PEARLM enhance recommendation in terms of serendipity, diversity, novelty and coverage?
\item What factors, such as the use of KGCD and direct embedding learning, as well as sampled path properties and model parameter size, influence the performance of PEARLM?

\end{enumerate}

\subsection{Experimental Setup}
\label{sec:exp-setup}
\noindent{\bf Datasets and Knowledge Graphs}
 We conducted experiments on two datasets, MovieLens1M (ML1M) \cite{ml1m} and LFM1M, which is a subset extracted by \cite{10.1007/978-3-031-28241-6_1} from LFM1B (LASTFM) \cite{lastfm-dataset} and employed in recent studies \cite{10.1145/3523227.3547374, 10.1145/3605357}. 
For ML1M, we adopted the KG originally generated in \cite{CaoWHHC19}. We adopted the KG generated in \cite{Wang00LC19} for LASTFM. We utilised the same datasets and KGs preprocessed in \cite{10.1007/978-3-031-28241-6_1} as well as their data preparation and split strategy. The stats of the employed dataset are reported for reference in Table~\ref{tab:data-stats}.

\noindent{\bf Data Preprocessing.} 
For all datasets, we follow \cite{10.1145/3523227.3547374} and remove products (and their interactions) absent in the KG as an entity. To control sparsity, we remove products and users that appear less than 5 times. 
Both KGs were pre-processed as in \cite{10.1145/3523227.3547374}, to make triplets uniformly formatted. 
Specifically, we only consider triplets composed of a product as the entity head and an external entity as the entity tail, so as to be able to analyse both knowledge-aware and path-based methods. Considering triplets having external entities or products as the head and tail entities would have required to craft additional meta-paths (needed for path-based methods) compared to the benchmarked models, going beyond the scope of our work. 

\noindent{\bf Data Preparation and Split.} 
For each dataset, we first sorted the interactions of each user chronologically. 
We then performed a training-validation-test split, following a time-based hold-out strategy, with the 60\% oldest interactions in the training set, the following 20\% for validation, and the 20\% most recent ones as the test set, which follows the setups of \cite{10.1145/3477495.3532041, 10.1007/978-3-031-28241-6_1}. 
We used these pre-processed datasets to train, optimise, and test each benchmarked model, facilitating an evaluation in a realistic setting, reflecting non-stationary, evolving interaction patterns over time.

\noindent{\bf Benchmarked models.} Our baselines\footnote{The detailed description of each baseline and the selected fine-tuned
hyper-parameters are listed in the README of our repository.}
included three non-explainable knowledge-aware models based on regularisation (CKE \cite{CKE10.1145/2939672.2939673}, CFKG \cite{AiACZ18}, KGAT \cite{Wang00LC19}), three knowledge-aware models based on reasoning paths (PGPR \cite{10.1145/3331184.3331203}, CAFE \cite{10.1145/3340531.3412038}, UCPR \cite{usercentric}), and the only proposed path-based causal language model (PLM\footnote{The original code of this method was not made available by the authors and our reproduced procedure can be found in the GitHub repository.}~\cite{10.1145/3485447.3511937}). We used the same framework of \cite{10.1007/978-3-031-28241-6_1}. Taking into account PEARLM, we report the best results gathered among a range of experiments conducted across the three dimensions: the language model size, the sampling size proportionate to the user set $|U|$, and the hop size (Section~\ref{sec:rq3}). For both PLM and PEARLM, we opt to train for a fixed number of iterations to emulate a realistic operational environment. This approach mirrors real-world scenarios where recommender systems are trained within a fixed computational budget or timeframe, which naturally restricts the number of training iterations feasible. This decision allows our results to reflect more practical performance expectations and constraints. For both models, at inference time we generated 100 sequences using 30 beams, divided into 5 beam groups with a diversity penalty of 0.3.

\begin{table}[!t]
\caption{Interaction and knowledge information.}
\label{tab:data-stats}
\centering
\setlength{\tabcolsep}{3pt}
\resizebox{1\linewidth}{!}{
\begin{tabular}{l|rr|rr|l|rr|rr}
\hline
& \multicolumn{2}{c|}{\textbf{ML1M}} & \multicolumn{2}{c}{\textbf{LFM1M}} & & \multicolumn{2}{c|}{\textbf{ML1M}} & \multicolumn{2}{c}{\textbf{LFM1M}} \\
\hline
\textbf{Interaction} & \multicolumn{2}{c|}{} & \multicolumn{2}{c}{} & \textbf{Knowledge} & \multicolumn{2}{c|}{} & \multicolumn{2}{c}{} \\
\hline
Users & \multicolumn{2}{c|}{6,040} & \multicolumn{2}{c}{4,817} & Entities (Types) & \multicolumn{2}{c|}{13,804 (12)} & \multicolumn{2}{c}{17,492 (5)} \\
Products & \multicolumn{2}{c|}{2,984} & \multicolumn{2}{c}{12,492} & Relations (Types) & \multicolumn{2}{c|}{193,089 (11)} & \multicolumn{2}{c}{219,084 (4)} \\
Interactions & \multicolumn{2}{c|}{932,295} & \multicolumn{2}{c}{1,091,275} & Sparsity & \multicolumn{2}{c|}{0.0060} & \multicolumn{2}{c}{0.0035} \\
Density & \multicolumn{2}{c|}{0.05} & \multicolumn{2}{c}{0.01} & Avg. Degree Overall & \multicolumn{2}{c|}{28.07} & \multicolumn{2}{c}{25.05}\\
& \multicolumn{2}{c|}{-} & \multicolumn{2}{c}{-} & Avg. Degree Products & \multicolumn{2}{c|}{64.86} & \multicolumn{2}{c}{17.53} \\

\hline

\end{tabular}}
\vspace{-4mm}
\end{table}

\noindent{\bf Evaluation Metrics}
We evaluated recommendation utility on top-10 recommended lists ($n=10$), using metrics like Normalized Discounted Cumulative Gain (NDCG) and Mean Reciprocal Rank (MRR). NDCG evaluates the quality of recommendations based on their ranking, giving higher weights to relevant products appearing earlier in the top-k list. MRR evaluates how well the model recommends the most relevant products at the top ranks. Furthermore, Precision and Recall metrics give us an overview of the model's accuracy and completeness. Precision calculates the proportion of recommended products that are relevant, while Recall identifies the proportion of relevant products that are recommended. In addition to these, we computed beyond-utility goals \cite{10.1145/2926720}, which assess the model's ability to provide diverse, unexpected, and novel recommendations. These include Serendipity (SER), Diversity (DIV), Coverage (COV), and Novelty. 

\begin{table*}[!t]
\caption{Combined Metric Scores for Recommendation Utility [RQ1] and Beyond Utility Goals [RQ2].}
\centering
\resizebox{\linewidth}{!}{
\begin{tabular}{l|l|l|l|l|l|l|l|l|l|l|l|l|l|l|l|l|l|l}
\hline
\multicolumn{1}{c}{\textbf{Method}} & \multicolumn{8}{c}{\textbf{ML1M}} & \multicolumn{8}{c}{\textbf{LFM1M}} \\
\hline
& \multicolumn{4}{c|}{\textbf{Utility Metrics}} & \multicolumn{4}{c|}{\textbf{Beyond Utility Metrics}} & \multicolumn{4}{c|}{\textbf{Utility Metrics}} & \multicolumn{4}{c}{\textbf{Beyond Utility Metrics}} \\
& NDCG $\uparrow$ & MRR $\uparrow$ & Precision $\uparrow$ & Recall $\uparrow$ & SER $\uparrow$ & DIV $\uparrow$ & NOV $\uparrow$ & COV $\uparrow$ & NDCG $\uparrow$ & MRR $\uparrow$ & Precision $\uparrow$ & Recall $\uparrow$ & SER $\uparrow$ & DIV $\uparrow$ & NOV $\uparrow$ & COV $\uparrow$ \\
\hline
\texttt{CKE} & 0.30 & 0.23 & \underline{0.11} & \underline{0.04} & 0.67 & 0.38 & \underline{0.92} & 0.30 & \underline{0.33} & \underline{0.27} & \underline{0.12} & \underline{0.03} & 0.92 & 0.48 & 0.86 & 0.23 \\
\texttt{CFKG} & 0.27 & 0.21 & 0.10 & 0.03 & 0.22 & 0.41 & \underline{0.92} & 0.03 & 0.13 & 0.10 & 0.03 & 0.01 & 0.54 & 0.39 & 0.84 & 0.03 \\
\texttt{KGAT} & \underline{0.31} & \underline{0.24} & \underline{0.11} & \underline{0.04} & 0.49 & 0.40 & \underline{0.92} & 0.13 & 0.30 & 0.24 & 0.11 & 0.01 & 0.69 & 0.19 & 0.87 & \underline{0.49} \\
\hline
\texttt{PGPR} & 0.28 & 0.21 & 0.08 & 0.03 & 0.77 & \underline{0.43} & \underline{0.92} & 0.43 & 0.18 & 0.14 & 0.04 & 0.01 & 0.84 & \underline{0.62} & 0.81 & 0.14 \\
\texttt{UCPR} & 0.26 & 0.19 & 0.07 & 0.03 & 0.78 & 0.42 & \textbf{0.93} & 0.52 & 0.32 & 0.26 & 0.10 & 0.03 & 0.98 & 0.56 & \underline{0.87} & 0.40 \\
\texttt{CAFE} & 0.21 & 0.15 & 0.06 & 0.03 & 0.77 & \textbf{0.45} & \textbf{0.93} & 0.26 & 0.14 & 0.09 & 0.04 & 0.01 & 0.82 & \textbf{0.63} & 0.86 & 0.03 \\
\hline
\texttt{PLM} & 0.27 & 0.18 & 0.07 & 0.03 & \underline{0.90} & \textbf{0.45} & \textbf{0.93} & \underline{0.64} & 0.28 & 0.19 & 0.08 & 0.02 & \textbf{0.98} & \textbf{0.63} & \underline{0.87} & 0.45 \\
\texttt{PEARLM} & \textbf{0.44} & \textbf{0.31} & \textbf{0.13} & \textbf{0.08} & \textbf{0.93} & \textbf{0.45} & \textbf{0.93} & \textbf{0.8} & \textbf{0.58} & \textbf{0.51} & \textbf{0.29} & \textbf{0.12} & \underline{0.97} & 0.55 & \textbf{0.88} & \textbf{0.78} \\
\bottomrule
\multicolumn{17}{l}{For each dataset: best result in \textbf{bold}, second-best result \underline{underlined}.}\tabularnewline
\end{tabular}
}
\label{tab:rq-combined}
\end{table*}

\subsection{Recommendation Utility Comparison (RQ1)}
\label{sec:rq1}
In our initial analysis, we investigated whether our proposed method, PEARLM, could significantly improve the recommendation utility, a central objective of any recommendation system. For this purpose, we primarily focused on Normalized Discounted Cumulative Gain (NDCG), though we also examined other commonly used utility metrics such as MRR, Precision, and Recall. These additional measures largely followed the trend set by the NDCG results.\\
Table~\ref{tab:rq-combined} presents the results of this evaluation, clearly highlighting the strong performance of PEARLM in terms of NDCG. For the ML1M dataset, PEARLM achieves an NDCG score of 0.44, a considerable 42\% improvement over the next best performer KGAT which scores 0.31. Likewise, on the LFM1M dataset, PEARLM records an NDCG score of 0.59, outperforming the second-best CKE by 78.7\%.

These improvements can be attributed to the embedding direct learning of PEARLM which, unlike PLM, is not constrained to precomputed knowledge graph embeddings. This freedom allows PEARLM to construct more nuanced embeddings, which translates to higher-quality recommendations in the downstream task. 

Our results suggest a deeper understanding of the distinct behaviors of different graph-based recommendation models may be emerging. Knowledge Graph Embeddings (KGE) typically draw insights from a node's immediate neighbors to derive their embeddings. In contrast, models like CKE and KGAT, which fall under the broader category of Graph Neural Networks (GNNs), appear to extend this scope by considering multiple neighbors based on the number of layers they operate on. By aggregating information across these layers, these models potentially assimilate data from a broader neighborhood, functioning in a manner reminiscent of Breadth-First Search (BFS). This approach seemingly provides a more localized understanding, focusing predominantly on the immediate surrounding context of a node.

PEARLM, along with similar Causal Language Models (CLMs), appears to diverge from this breadth-oriented approach. The strategy employed by these models is akin to a Depth-First Search (DFS), where exploration aims to delve deeper into the graph to seek paths leading to distant nodes. This depth-focused traversal might provide a more expansive understanding of the graph, potentially allowing the model to integrate knowledge from both proximate and distant entities. The resulting embeddings are hypothesized to be richer and more detailed, capturing intricate interrelations and a holistic representation of user-product dynamics. These characteristics are believed to translate into superior recommendation outcomes.

\hlbox{Observation 1}{PEARLM's depth-centric exploration approach crafts detailed embeddings, capturing intricate graph relationships. This results in significantly enhanced performance in the recommendation downstream task compared to neighbour-focused approaches.}


\subsection{Beyond-Utility Comparison (RQ2)}
\label{sec:rq2}
In this subsection, we delve into the beyond-accuracy performance of the compared models, evaluating them against metrics of Coverage, Serendipity, Diversity, and Novelty.
Table~\ref{tab:rq-combined} showcases PEARLM's noteworthy performance. On the ML1M dataset, PEARLM achieves a gain of 3.33\% in Serendipity compared to the second-best model (UCPR) and 19.2\% over the third-best (PLM). For Diversity and Novelty, PEARLM's metrics are aligned with the top-performing models. In terms of Coverage, the differences range from 25\% to 53\% when contrasted with the second-best (PLM) and the third-best competitors (UCPR, KGAT), respectively. On the LFM1M dataset, PEARLM exhibits a slight decrease of -1\% in Serendipity compared to PLM but gains 6.5\% over CKE. A decrease of 6.35\% in Diversity is noted against the top performer (PLM). However, for Coverage, PEARLM shows a 73\% improvement over PLM and 47\% over CKE, while Novelty remains competitive with the leading methods.

While PEARLM surpasses the majority of methods across various metrics, it is important to note that it does not universally outperform all models. In certain metrics, other models, notably those from the same family as PEARLM, such as PLM, demonstrate slightly better results. This underscores the intricate trade-offs involved in optimising for multiple evaluation criteria and further emphasises the nuanced complexity of recommendation tasks \cite{10.1145/3523227.3547425}.

The analysis of the results also reveals the distinct strengths of path reasoning methods and knowledge-aware models. As observed in previous studies \cite{10.1007/978-3-031-28241-6_1}, path reasoning approaches like PGPR, UCPR, and CAFE tend to excel in Serendipity and Diversity, underscoring their ability to provide novel and varied recommendations. Conversely, knowledge-aware models such as CKE, CFKG, and KGAT generally perform better in Coverage and Novelty, suggesting their adeptness at tapping into a wide array of products and recommending less popular, yet relevant, options.

The results highlight the potential of path causal language models like PLM and PEARLM. They seamlessly integrate a diverse array of recommendations—varied, unexpected, and broad—while ensuring explainability through the provision of reasoning paths.

\hlbox{Observation 2} {PEARLM's results reveal an integration of the best attributes from both path reasoning and knowledge-aware models. It consistently delivers performance that is either superior or, at the very least, comparable to top-performing models across key metrics.}

\subsection{Utility-Oriented Ablation Study (RQ3)} \label{sec:rq3}
In our final analysis, we explore the critical factors that significantly influence the performance of our Path Language Modeling (PLM) framework. These factors include the properties of sampled paths (e.g., sample size and path length), model parameter size, and the utilization of KG Constraint Decoding (KGCD). We also assess the impact of direct embedding learning within the model architecture.

\begin{table}[!t]
\caption{NDCG and MRR values with different model embeddings configurations TransE initialisation vs Direct learning used to initialise PEARLM and and PLM.}
\vspace{-2mm}
\resizebox{0.9\linewidth}{!}{
\begin{tabular}{l|l|l|l|l|l|l|}
\label{tab:ablation-embedding}
\textbf{Architecture} & \multicolumn{2}{c}{\textbf{ML1M}} & \multicolumn{2}{c}{\textbf{LFM1M}} \\
\hline
& NDCG & MRR & NDCG & MRR  \\
\hline
PLM w/ TransE& 0.27 & 0.18 & 0.28 & 0.19  \\
PLM w/ direct learning & \underline{0.40} & \underline{0.28} & \underline{0.51} & \underline{0.47} \\
PEARLM w/ TransE &  0.29 & 0.20 & 0.3 & 0.21\\
PEARLM w/ direct learning & \textbf{0.44} & \textbf{0.31} & \textbf{0.59} & \textbf{0.51} \\
    \bottomrule
\end{tabular}
}
\vspace{-4mm}
\end{table}

\noindent{\bf Impact of Direct Embedding Learning}
In this initial analysis, we aim to understand the impact of learning a new set of embeddings specifically for the path language modeling task instead of initializing them with knowledge graph embeddings as proposed in the original PLM framework. Table~\ref{tab:ablation-embedding} presents results obtained by training PLM and PEARLM with different embedding initializations. Both models used a sample size of 250 and distilgpt2 as the base, inferring their recommendations using KGCD to reduce the combinatory number of experiments. The results highlight that direct learning profoundly impacts the model's ability to produce relevant recommendation products and is the principal reason for PEARLM's significant gains compared to the state-of-the-art.

\begin{table}[!t]
\caption{NDCG and PFR@K=3 values with and without constraint decoding for PEARLM and PLM.}
\vspace{-2mm}
\resizebox{0.9\linewidth}{!}{
\begin{tabular}{l|l|l|l|l|l|l|}
\label{tab:ablation-gcd}
\textbf{Architecture} & \multicolumn{2}{c}{\textbf{ML1M}} & \multicolumn{2}{c}{\textbf{LFM1M}} \\
\hline
& NDCG & PFR@3 & NDCG & PFR@3  \\
\hline
PLM w/o KGCD & 0.27 & 0.06 & 0.28 & 0.1  \\
PLM w/ KGCD & 0.27 & \textbf{1} & 0.28 & \textbf{1} \\
PEARLM w/o KGCD & \textbf{0.45} & 0.11 & \textbf{0.60} & 0.14\\
PEARLM w/ KGCD & \underline{0.44} & \textbf{1} & \underline{0.59} & \textbf{1} \\
    \bottomrule
\end{tabular}
}
\end{table}

\begin{table}[!t]
\caption{NDCG values with different model architectures configurations (distilgpt2, gpt2, gpt2-large) used to initialise PEARLM and sample size (250, 500, 1000) across different path lengths (3 hops).}
\vspace{-2mm}
\resizebox{0.9\linewidth}{!}{
\begin{tabular}{l|l|l|l|l|l|l}
\textbf{Architecture} & \multicolumn{3}{c}{\textbf{ML1M}} & \multicolumn{3}{c}{\textbf{LFM1M}} \\
\hline
                   & \multicolumn{6}{c}{Sample Size (hop=3)}      \\
& 250 & 500 & 1000 & 250 & 500 & 1000 \\
\hline
PLM & 0.26 & 0.27 & 0.26 & 0.27 & 0.27 & 0.28 \\
distilPEARLM & \underline{0.43} & \underline{0.43} & \textbf{0.44} & 0.55 & \textbf{0.58} & \textbf{0.58} \\
PEARLM-medium & \underline{0.43} & \underline{0.43} & 0.42 & 0.51 & 0.56 & \underline{0.57} \\
PEARLM-large & 0.41 & 0.42 & 0.41 & 0.52 & 0.56 & \textbf{0.58} \\
    \bottomrule
\end{tabular}
}
\vspace{-4mm}
\resizebox{0.9\linewidth}{!}{
\begin{tabular}{l|l|l|l|l|l|l}
\textbf{Architecture} & \multicolumn{3}{c}{\textbf{ML1M}} & \multicolumn{3}{c}{\textbf{LFM1M}} \\
\hline
                   & \multicolumn{6}{c}{Sample Size (hop=5)}      \\
& 250 & 500 & 1000 & 250 & 500 & 1000 \\
\hline
PLM & 0.25 & 0.25 & 0.25 & 0.25 & 0.25 & 0.25 \\
distilPEARLM & \textbf{0.40} & 0.38 & 0.37 & 0.49 & \underline{0.53} & \textbf{0.54} \\
PEARLM-medium & 0.37 & 0.38 & 0.38 & 0.45 & 0.51 & \underline{0.53} \\
PEARLM-large & 0.38 & \underline{0.39} & 0.37 & 0.43 & 0.51 & \underline{0.53} \\
    \bottomrule
\end{tabular}
}
\label{tab:rq3-utility-by-sample}
\end{table}

\noindent{\bf Impact of Graph Constraint Decoding.}
In a second analysis, we explore the impact of KG Constraint Decoding (KGCD) and its transferability, applying it to PLM and omitting it in PEARLM for utility measurements, using NDCG as a proxy. We measure PFR@K at maximum path lengths for all benchmarked models, demonstrating that utilizing KGCD results in a PFR@K of 1 Table~\ref{tab:ablation-gcd} shows that although omitting KGCD can slightly enhance the utility of the produced recommendations, its benefits in terms of path faithfulness are significant. Generally, a smaller value of PFR@K can only be achieved with KGCD if, for a given user, there are fewer than \(k\) top-\(k\) size paths connecting the user to unseen products. This highlights that KGCD may be limited by larger top-\(k\) sizes or when a user is not connected to a sufficient number of items.

\noindent{\bf Impact of Model and Sample Size.}
Table~\ref{tab:rq3-utility-by-sample} elucidates the intricate relationship between model architecture and sample size, contrasting the effects of 3-hop and 5-hop paths across different configurations of PEARLM, based on GPT2 architecture variants (distil, medium, large). Prior research \cite{wang2018explainable} shows that while longer paths capture a richer diversity of entity relationships, they may compromise interpretability due to increased complexity and noise.

For the 3-hop paths on the ML1M dataset, distilPEARLM achieves optimal NDCGs at larger sample sizes. However, the PEARLM-large model exhibits marginally lesser performance, suggesting potential diminishing returns with increased model size. A similar pattern is observed on the LFM1M dataset, where distilPEARLM consistently outperforms other models, achieving peak NDCGs at sample sizes of 500 and 1000. Considering the 5-hop paths, our results indicate diminished performance regardless of the sample size and model size adopted. This may be caused by the fact that these paths contain additional noise, which reduces the model's ability to generalize.

It is crucial to note that all architectures were trained with the same number of iterations, which may not adequately favor larger models or datasets that require more extensive training epochs to converge. Further analysis may reveal that larger models and increased sample sizes slightly enhance NDCG, aligning with the scaling laws of transformers \cite{kaplan2020scaling}.

\hlbox{Observation 3} {Direct Embedding Learning drastically improves recommendation accuracy and outshines traditional methods. KGCD enhances the faithfulness of paths, crucial for reliable outputs, while the interaction between model size and sample size highlights that shorter paths better align with user history, balancing performance with computational efficiency.}

\section{Conclusion}
In this work, we introduce PEARLM, a novel framework for explainable recommendations based on causal language modeling with paths in knowledge graphs. A critical advancement in PEARLM is its response to the issue of path hallucination, which was identified through a newly defined metric that measures the prevalence of corrupted paths. This metric provides a quantitative foundation to evaluate and understand the impact of inaccuracies in generated paths. Furthermore, we conducted a user survey to explore the negative effects of these inaccuracies on user trust and satisfaction. The survey results confirmed that hallucinated paths significantly undermine user trust, highlighting the necessity for more accurate and verifiable path generation methods like PEARLM. By integrating Knowledge Graph-Constrained Decoding (KGCD), PEARLM ensures that all generated paths are authentic and representative of true KG relationships, thereby eliminating the risk of misleading users and allowing this model to be safely deployed in real-world scenarios. Although KGCD has been proposed for decoding in PEARLM, it is important to note that this constrained decoding technique, which ensures the faithfulness of explanations, is model-agnostic. Hence, this approach can be adopted by any path-based language model to ensure the reliability of generated explanations. We demonstrated its transferability by applying it to existing models. The model's performance, however, depends on the integrity and comprehensiveness of the underlying knowledge graph, pinpointing an essential direction for future research in data enhancement and graph augmentation techniques.

Moreover, PEARLM outperforms existing methods in recommendation utility and beyond by employing direct learning of token embeddings (KG entities and relations), which has been empirically shown to be the main component affecting its performance. Furthermore, our findings suggest that even with a reduced number of paths sampled as training data and smaller model sizes, there is minimal loss in efficacy, indicating that it can be deployed effectively even in resource-constrained real-world applications.

Future work will explore the scalability of PEARLM to larger model architectures and its generalizability across various graph-oriented tasks, such as link prediction and node classification. Additionally, we plan to refine the explanation generation mechanism, integrating sophisticated path generation with versatile template creation to accommodate diverse user preferences and use cases.

\balance
\bibliographystyle{ACM-Reference-Format}
\bibliography{sn-bibliography}

\end{document}